\documentclass[pre,superscriptaddress,twocolumn,amsmath,amssymb,floatfix,showpacs]{revtex4}

\usepackage{graphicx}
\usepackage{dcolumn}
\usepackage{bm}

\begin{document}


\title{Universality classes and crossover behaviors in non-Abelian
directed sandpiles}
\author{Hang-Hyun Jo}
\altaffiliation{Present address: BECS, School of Science and Technology, Aalto University, P.O. Box 12200, FI-00076}
\affiliation{School of Physics, Korea Institute for Advanced
Study, Seoul 130-722, Korea}

\author{Meesoon Ha}
\email[Corresponding author: ]{msha@kaist.ac.kr}
\affiliation{Department of Physics, Korea Advanced Institute of
Science and Technology, Daejeon 305-701, Korea}
\affiliation{School of Physics, Korea Institute for Advanced
Study, Seoul 130-722, Korea}

\date{\today}

\begin{abstract}
We study universality classes and crossover behaviors in
non-Abelian directed sandpile models, in terms of the metastable
pattern analysis. The non-Abelian property induces spatially
correlated metastable patterns, characterized by the algebraic
decay of the grain density along the propagation direction of an
avalanche. Crossover scaling behaviors are observed in the grain
density due to the interplay between the toppling randomness and
the parity of the threshold value. In the presence of such
crossovers, we show that the broadness of the grain distribution
plays a crucial role in resolving the ambiguity of the
universality class. Finally, we claim that the metastable pattern
analysis is important as much as the conventional analysis of
avalanche dynamics.

\end{abstract}

\pacs{05.70.Ln, 05.65.+b, 64.60.Ht}




\maketitle

\section{Introduction}
\label{Sect:intro}

Scale invariance in avalanche systems is ubiquitously observed in
nature, such as earthquakes, sandpiles, and Barkhausen avalanches
of ferromagnetic materials. The statistics of avalanches follows
some universal power-law
distribution~\cite{CM2005,Newman2005,UMM1995}. Metastable patterns
between avalanches also exhibit spatially long-range correlations.
A fractal structure in the crust of the earth formed by seismic
events is one of such examples. It is, however, still questionable
to find generic mechanisms for avalanche systems that can explain
both the statistics of avalanches and long-range correlations in
metastable patterns.

To reveal the underlying common mechanism of scale invariance in
avalanche systems, Bak, Tang, and Wiesenfeld (BTW) first proposed
the paradigm of self-organized criticality with an Abelian
deterministic version of undirected sandpile models. Since then,
lots of BTW variants have been suggested and
studied~\cite{Bak1987,Jensen1998,Dhar2006}. In modeling
sandpiles, the balance between slow driving and dissipation is the
key ingredient. Grains are slowly added, toppled instantly
whenever the instability threshold is overcome, and finally
dissipated at boundaries of the system. It has been tested whether
the universality class of avalanche dynamics can be changed by the
breaking of Abelian symmetry~\cite{YCZhang1989} or the
consideration of stochasticity~\cite{Manna1991} in toppling rules.
The issue is still controversial due to conflicting numerical
results~\cite{Ben-Hur1996,Lubeck1997a,Lubeck1997b,Milshtein1998,Dickman2003}.
The Abelian symmetry here means that the order of toppling events
does not affect the final state.

Contrary to undirected sandpile models, directed sandpile models
(DSMs) with a preferred direction of avalanche propagation turn
out to be more tractable analytically as long as they have the
Abelian
symmetry~\cite{Dhar1989,Pastor-Satorras2000,Paczuski2000,Kloster2001}.
This is because the Abelian symmetry in DSMs lets metastable grain
patterns be fully uncorrelated. However, most real avalanche
systems often exhibit spatially correlated grain patterns. It
means that Abelian DSMs are not suitable enough to describe such
systems and non-Abelian DSMs are more natural to be considered.

When the Abelian symmetry is broken in
DSMs~\cite{Hughes2002,Pan2005,JoHa2008}, spatially long-range
correlations emerge in metastable grain and scar patterns. Scars
are defined as traces of avalanche boundaries. Both scar and grain
densities decay algebraically along the preferred direction of
avalanche propagation with the same decay exponent. In our earlier
work~\cite{JoHa2008}, additional scaling relations are derived
from avalanche flow equations. All avalanche exponents can be
written in terms of the grain density exponent. With the
stochastic toppling rule, the broken Abelian symmetry seems not to
change the universality class. However, it changes the scaling
behavior of metastable patterns with the non-zero value of the
grain density exponent. When the toppling rule becomes
deterministic, the Abelian symmetry turns out to be relevant to
the universality class and non-Abelian DSMs belong to the
mean-field (MF) universality class. These results were numerically
confirmed in the simplest models~\cite{JoHa2008}.

However, the simplest discrete DSM with the non-Abelian
deterministic toppling rule cannot be unique due to the interplay
between the toppling rule and the lattice structure. For an
example, if the number of grains at an unstable site is $3$ and
the number of its neighboring sites is $2$, there are several ways
to topple the last grain into one of the neighboring sites. Two
versions of the toppling bias for the last grain were considered,
which lead to some deviation from MF results~\cite{JoHa2008}.

In this paper, we check out the validity of universality classes
in non-Abelian DSMs against the change of the toppling bias for
the last grain and the instability threshold value. We also report
various crossover behaviors between universality classes. One
might expect that the toppling bias effect becomes negligible as
the threshold value increases. It turns out to be only true for
the fully biased case. For alternatively and/or partially biased
cases, the parity of the threshold value plays a crucial role in
avalanche dynamics. If the threshold value is odd, the
stochasticity in toppling rules becomes relevant. As a result, the
{\it deterministic} models belong to the non-Abelian {\it
stochastic} universality class. It is somehow surprising because
one believes the threshold value in sandpile models to be
irrelevant to universality classes, except for crossover behaviors
discussed in undirected cases by L\"{u}beck~\cite{Lubeck2000b}.

This paper is organized as follows. In Sec.~\ref{Sect:model}, we
briefly describe various DSMs including those proposed
in~\cite{JoHa2008}. In Sec.~\ref{Sect:flow analysis}, we discuss
the role of the threshold value and the toppling bias in
determining universality classes by the conventional analysis of
avalanche dynamics and the metastable pattern analysis. We also
argue possible scenarios for universality classes and crossover
behaviors, which are numerically confirmed in
Sec.~\ref{Sect:numerics}. Finally, we conclude this paper in
Sec.~\ref{Sect:summary} with the summary of main results and some
suggestions.

\section{Directed Sandpile Models}
\label{Sect:model}

\begin{figure}[b]
\includegraphics[width=0.6\columnwidth,angle=0]{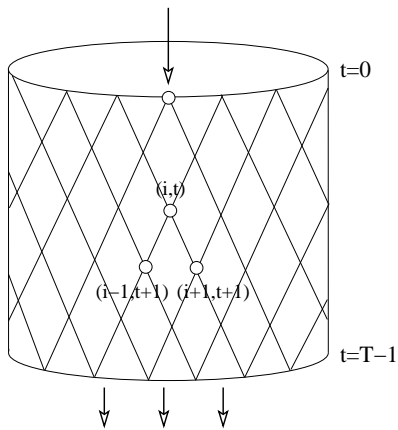}
\caption{A two-dimensional $\frac{\pi}{4}$-rotated lattice for
DSMs.} \label{fig:lattice}
\end{figure}

We consider DSMs defined on a two-dimensional
$\frac{\pi}{4}$-rotated lattice of sizes $(L_\perp,L_\|)$. For
convenience, we call $L_\perp$ and $L_\|$ as $L$ and $T$,
respectively. The preferred direction of avalanche propagation is
denoted by the ``layer'' $t=0,1,...,T-1$ with open boundary
conditions. The transverse direction is denoted by
$i=0,1,\cdots,L-1$ with periodic boundary conditions (see
Fig.~\ref{fig:lattice}). Initially, to each site of the lattice an
integer value (the number of grains), $z(i,t) \in
\{0,1,...,z_c-1\}$, is assigned. Here $z_c$ denotes the
threshold value.

Given a stable configuration where $z(i,t)<z_c$ for all sites
$(i,t)$, one grain is added at a randomly chosen site on the top
layer, $z(i,0)\rightarrow z(i,0)+1$. For any unstable site with
$z(i,t)\geq z_c$, including that on the top layer, grains topple
down to its left and right nearest-neighboring sites on the next
layer $t+1$.
\begin{eqnarray}
z(i,t)&\to& z(i,t)-\Delta_{i,t},\nonumber\\
z(i\pm 1,t+1)&\to& z(i\pm 1,t+1)+\tilde\Delta_{i\pm 1,t+1}.
\end{eqnarray}
Here $\Delta_{i,t}$ represents the number of grains outgoing from
$(i,t)$ and $\tilde\Delta_{i\pm1,t+1}$ does the number of grains
incoming to $(i\pm 1,t+1)$ from $(i,t)$. The number of toppled
grains is locally conserved: $\Delta_{i,t}=
\tilde\Delta_{i-1,t+1}+ \tilde\Delta_{i+1,t+1}$.

The toppling at one layer may cause another toppling on the next
layer. At unstable sites on the bottom layer $t=T-1$, toppled
grains are dissipated out of the system. Only after another stable
configuration is recovered by a series of toppling events,
denoting an avalanche, a new grain is added on the top layer,
$t=0$, to keep generating another avalanche.
\begin{figure}[t]
\includegraphics[width=0.8\columnwidth,angle=0]{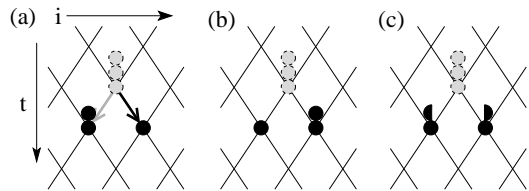}
\caption{Toppling rules of non-Abelian deterministic DSMs on a
two-dimensional tilted lattice~\cite{JoHa2008}: (a)~aND, (b)~bND,
and (c)~cND. The gray colored grains and arrow represent the state
before toppling and the black colored ones represent the state
after toppling.} \label{fig:toppling}
\end{figure}

\begin{figure*}[t]
\includegraphics[width=1.6\columnwidth,angle=0]{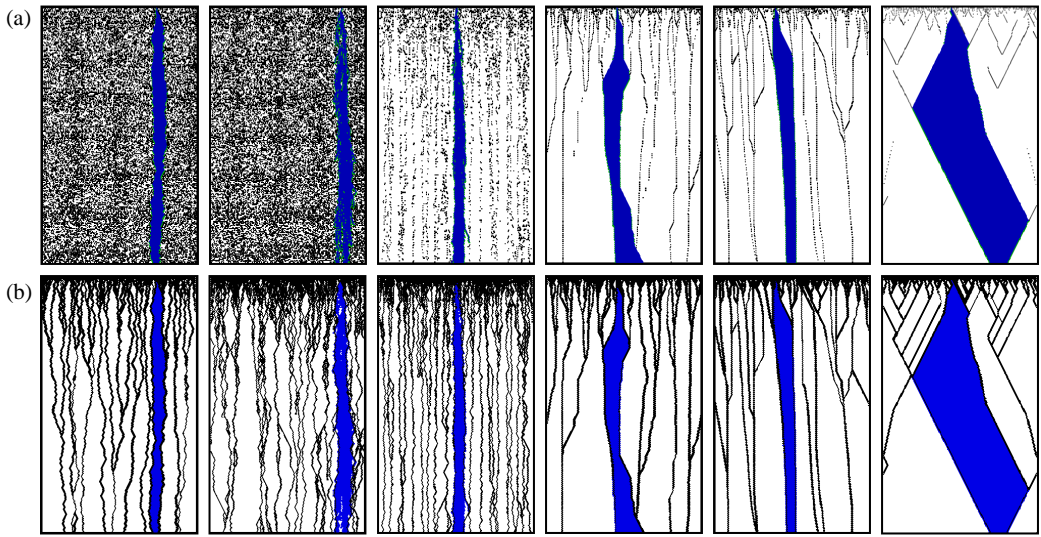}
\caption{(Color online) Typical metastable grain (a) patterns and
(b) scars of six DSMs (from left to right, AD, AS, NS, aND, bND,
and cND) at $z_c=2$: occupied sites and scars are
shown as black dots and the typical shapes of dissipative
avalanches consisting of toppled sites as blue (shaded) areas on a
lattice with $L=150$ (horizontally) and $T=250$ (vertically).}
\label{fig:patterns_vs_scars}
\end{figure*}

By setting the toppling matrices, $\Delta$ and $\tilde\Delta$, one
may consider several variants of DSMs. Without losing any
generality, we set $\Delta_{i,t}=z_c$ in the Abelian version and
$\Delta_{i,t}=z(i,t)\ge z_c$ in the non-Abelian version. The
values of $\tilde\Delta_{i\pm 1,t+1}$ can be either stochastically
or deterministically determined. For the stochastic case, each
grain at the unstable site is toppled at random to one of the
nearest neighbors at the next layer. Abelian and non-Abelian
stochastic DSMs are denoted as AS and NS for short, respectively.
For Abelian deterministic DSMs (AD for short), one should fix the
values of $\tilde\Delta_{i\pm 1,t+1}$ under the condition
$\tilde\Delta_{i-1,t+1}+\tilde\Delta_{i+1,t+1}=z_c$. For an
example, each value of $\tilde\Delta_{i\pm 1,t+1}$ is set as one
half of $z_c$ when $z_c$ is even.

For non-Abelian deterministic DSMs (ND for short) where the
simplest setup is not clear, the following three versions are
considered in~\cite{JoHa2008}:
\begin{eqnarray}\label{eq:NDrules}
{\rm (i)}~\tilde\Delta_{i\pm 1,t+1} &=&\left\{\begin{array}{ll}k &
\textrm{if $z(i,t)=2k$}\\
k+\delta_{i\pm 1, a(i,t)} & \textrm{if
$z(i,t)=2k+1$},\end{array}\right. \nonumber\\
{\rm (ii)}~\tilde\Delta_{i\pm 1,t+1}&=&\left\{\begin{array}{ll}k &
\textrm{if $z(i,t)=2k$}\\
k+\delta_{i\pm 1, i+1} & \textrm{if
$z(i,t)=2k+1$},\end{array}\right. \nonumber\\
{\rm (iii)}~\tilde\Delta_{i\pm 1,t+1} &=&z(i,t)/2.
\end{eqnarray}
Here, $k$ is a positive integer and $\delta_{ij}$ denotes the
Kronecker delta function. We call (i) the alternative bias version
(aND)~\cite{Pan2005}, (ii) the full bias version (bND), and (iii)
the continuous version without bias (cND). For the
aND, a ``toppling arrow'' $a(i,t)$ of each site initially points
to one of its neighbors, say, $i-1$, in Fig.~\ref{fig:toppling}(a).
Whenever each grain is toppled at the unstable site, the direction
of the arrow flips to the other neighbor.
Figure~\ref{fig:toppling} shows the case of $z(i,t)=3$ and
$z_c=2$. It is worthwhile to note that any toppling bias here is
applied only to the last grain when the number of toppled grains
at the unstable site, $z(i,t)$, is odd.

In order to check the validity of earlier ND versions, we
introduce another ND version that covers both the aND and the bND
by controlling the bias of the last grain at the toppled site as
follows:
\begin{equation}\label{eq:pNDrule}
{\rm (iv)}~\tilde\Delta_{i\pm 1,t+1} =\left\{\begin{array}{ll}k &
\textrm{if $z(i,t)=2k$}\\
k+\delta_{i\pm 1, r(i,t)} &\textrm{if
$z(i,t)=2k+1$}.\end{array}\right.
\end{equation}
Here $r(i,t)$ takes $i+1$ with a probability $p$ or $i-1$ with a
probability $1-p$ for each toppling event. For convenience, only
the range of $\frac{1}{2}\leq p\leq 1$ is considered. We call (iv)
the partial bias version (pND for short). This model is exactly the
same as the bND when $p=1$. Moreover, it is almost the same as the
aND when $p=\frac{1}{2}$, except that toppling arrows are
initially quenched in the aND or completely annealed in the pND.

An avalanche of DSMs can be quantified as mass $s$ (the total
number of toppled grains), duration $t$ (the number of affected
layers), area $a$ (the total number of distinct toppled sites),
width $w$ (the distance between avalanche boundaries), and height
$h$ (the number of toppled grains per toppled site). There is no
characteristic scale in avalanche dynamics, except for $T$ as long
as $L$ is sufficiently larger than the maximum width of the
avalanche. It is well known that all probability distribution
functions of avalanche quantities are expected to follow the
simple scaling form as
\begin{equation}
P(x,T)\sim x^{-\tau_x} f\left( x/T^{D_x} \right)
\end{equation}
for $x \in \{s,t,a,w,h\}$. Hence, any two quantities $x$ and
$y$ scale, using the conditional expectation value, as
\begin{equation}
E[y|x] \sim x^{\gamma_{yx}}.
\end{equation}
The relative exponent $\gamma_{yx}=(\tau_x-1)/
(\tau_y-1)={D_y}/{D_x}$ is derived from the identity
$P(x,T)dx=P(y,T)dy$.

We take full advantage of the relations, $D_t=1$ and $\langle
s\rangle\sim T$ in DSMs, where $\langle\cdot\rangle$ denotes the
ensemble average. Based on the reasonable assumption for the
compactness of the avalanche in low dimensional systems, i.e.,
$a\sim wt$ and $s\sim ah$, we obtain the following scaling
relations:
\begin{eqnarray}
\gamma_{xt}=D_x\ \textrm{for any}\ x,\nonumber\\
D_s(2-\tau_s)=1,\\
D_a=D_w+1,\ \textrm{and}\ D_s=D_a+D_h,\nonumber
\end{eqnarray}
which implies only two independent exponents left.

Concerning metastable patterns of DSMs, we are able to define two
additional scaling exponents. In order to quantify metastable
patterns, we denote the probability to find $z$ remaining grains
at each layer $t$  as
\begin{equation}\label{eq:Qzt}
Q(z,t)\equiv\Pr[z(i,t)=z],
\end{equation}
where $\sum_{z=0}^{z_c-1}Q(z,t)=1$. Along the avalanche
propagation direction, one can measure the grain density as
\begin{equation}
\rho(t)\equiv \frac{1}{L} \sum_{i=0}^{L-1} \langle z(i,t)\rangle
=\sum_{z=0}^{z_c-1} zQ(z,t),
\end{equation}
where $\langle\cdot\rangle$ denotes the recurrent configuration
average. Similarly, the fraction of occupied sites at each layer,
namely, the grain occupation density, can be measured as
\begin{equation}
\rho_{_\theta}(t)\equiv \frac{1}{L} \sum_{i=0}^{L-1} \langle
\theta[z(i,t)]\rangle =\sum_{z=1}^{z_c-1} Q(z,t),
\end{equation}
where $\theta(z)$ denotes the Heaviside step function that gives
$0$ for $z=0$ and $1$ for $z>0$. Both $\rho(t)$ and
$\rho_{_\theta}(t)$ decay algebraically with the grain density
exponent $\alpha$ at large $t$ in non-Abelian DSMs as
\begin{equation}
\rho(t)\sim\rho_{_\theta}(t)\sim t^{-\alpha}.
\end{equation}
For Abelian DSMs with an uniform grain density ($\alpha=0$), we
consider the scar density $\rho_{\rm{sc}}(t)$ and the scar
exponent $\alpha_{\rm{sc}}$ with the same definition of $\rho(t)$
where $z(i,t)$ is just replaced by the quantity, $b(i,t)$. The
value of $b(i,t)$ is $1$ if the site $(i,t)$ has recently been a
part of avalanche boundaries, or $0$ otherwise. We call
$\{b(i,t)\}$ a scar.

\section{Avalanche dynamics}
\label{Sect:flow analysis}

We begin with the discussion of statistical properties observed in
metastable grain and scar patterns. The metastable pattern
analysis is important as much as the conventional analysis of
avalanche dynamics. As reported in~\cite{JoHa2008}, the scar
exponent is a good indicator for the universality classes of DSMs
since it is directly related to the exponent of the avalanche
width as $\alpha_{\rm{sc}}=D_w$. It is because the scar density is
inversely proportional to the typical avalanche width.
\begin{equation}
\rho_{\rm sc}(t)=\frac{1}{L} \sum_{i=0}^{L-1} \langle
b(i,t)\rangle\approx \frac{1}{L}\frac{L}{w(t)}=w(t)^{-1}.
\end{equation}
It is worthwhile to note that grains in non-Abelian DSMs can remain
only at avalanche boundaries, which implies $\alpha=\alpha_{\rm
sc}$. On the other hand, Abelian DSMs exhibit an uncorrelated and
uniform grain density in metastable grain patterns, i.e.,
$\alpha=0$, irrespective of the $\alpha_{\rm sc}$ value.

We intuitively argue the interplay between avalanche flow and
metastable grain or scar patterns in DSMs. Let us define $N(t)$ as
the number of grains toppled from the layer $t$ to the next layer
$t+1$ within an avalanche. It scales as $N(t)\sim w(t)h(t)\sim
t^{D_w+D_h}$ with the assumption for the compactness of an
avalanche. The evolution of $N(t)$, avalanche flow, can be written
as
\begin{equation}
\frac{dN(t)}{dt} \approx N(t)-N(t-1)=\sum_{i\in w(t)} n(i,t),
\label{eq:floweq}
\end{equation}
where $n(i,t)$ denotes the amount of avalanche flow at the site
$(i,t)$. The value of $n(i,t)$ depends on both the value of
$z(i,t)$ before toppling and the number of grains toppled from the
previous layer to the site $(i,t)$. The summation is over
avalanche boundaries and bulk sites, belonging to $w(t)$. Here
avalanche boundaries denote the outermost sites of an avalanche at
a given layer and avalanche bulk sites do the sites between
avalanche boundaries.

\subsection{Abelian cases}

In AD DSMs, it is well known that
metastable grain patterns are fully uncorrelated and that
probabilities to find $z(i,t)\in\{0,1,...,z_c-1\}$ grains at a
site are the same as $1/{z_c}$~\cite{Dhar1989}. The number
of toppled grains at every unstable site is $z_c$, such that
$D_h=0$. Since $n(i,t)=0$ at avalanche bulk sites, only avalanche
boundaries contribute to avalanche flow. At the right avalanche
boundary, the avalanche width $w$ increases by one if the toppling
at the rightmost site of the avalanche, say $(i,t)$, leads to a
toppling at a site $(i+1,t+1)$. The probability is given by
\begin{equation*}
\Pr[z(i+1,t+1)+\tilde\Delta_{i+1,t+1}\geq z_c]=
\frac{\tilde\Delta_{i+1,t+1}}{z_c}\equiv p_{_{\rm AD}}.
\end{equation*}
The probability to decrease $w$ by 1 is $1-p_{_{\rm AD}}$.
Likewise, probabilities to increase and decrease $w$ by one at the
left avalanche boundary are $1-p_{_{\rm AD}}$ and $p_{_{\rm AD}}$,
respectively. For the range of $0<p_{_{\rm AD}}<1$, avalanche
boundaries in the AD are exactly mapped onto the spatiotemporal
trajectories of random walks starting at the same site. The
avalanche flow equation is therefore described as
\begin{equation}
\frac{dN(t)}{dt}\approx\eta(t),
\end{equation}
where $\eta(t)$ denotes an uncorrelated noise with zero mean and
unit variance. As a result, we obtain $\alpha_{\rm
sc}=D_w=\frac{1}{2}$ with $D_h=0$, which yields all avalanche
exponent values. We note that the result in the AD is robust,
independent of the value of $z_c$ and any possible toppling bias
in the case of $p_{_{\rm AD}}\neq \frac{1}{2}$.

In AS DSMs, the fluctuations of
$\tilde\Delta_{i\pm 1,t+1}$ resulting from the stochastic toppling
rule enable multiple topplings (more than one toppling at the same
site). Metastable grain patterns are still fully uncorrelated
because of the characteristic of the Abelian symmetry
($\alpha=0$)~\cite{Paczuski2000,Kloster2001}. In that sense,
avalanche boundaries in the AS are also mapped onto the
trajectories of random walks, $\alpha_{\rm sc}=D_w=\frac{1}{2}$.
However, avalanche flow in the AS is mainly affected by avalanche
bulk sites. In general, the numbers of grains that bulk sites
receive from the previous layer can be written as $mz_c+l$, where
$m$ is a non-negative integer and $0\leq l\leq z_c-1$. If
$z(i,t)+l<z_c$ with the value of $z(i,t)$ in the metastable state,
$l$ grains would be left behind avalanche flow, i.e., $n(i,t)=-l$.
If $z(i,t)+l\geq z_c$, avalanche flow would sweep $z_c-l$ grains,
i.e., $n(i,t)=z_c-l$. Based on the reasonable assumption that the
probability distribution of $l$ is uniform, $n(i,t)$ is considered
as a random variable $\eta(i,t)$ with zero mean and finite
variance. As a result, the avalanche flow equation in the AS is
described as
\begin{equation}
\frac{dN(t)}{dt}\approx\sum_{i\in w(t)}
\eta(i,t)\approx\sqrt{w(t)}\eta(t),\label{eq:floweqAS}
\end{equation}
which gives $D_h=\frac{1}{4}$. A positive value of $D_h$ reflects
the existence of the multiple toppling. Once again, above results
are also robust, independent of the value of $z_c$. The toppling
bias can be considered as setting the probabilities of each grain
toppled to the right and the left nearest neighbors at the next
layer with $p_{_{\rm AS}}$ and $1-p_{_{\rm AS}}$, respectively. In
other words,
$$
p_{_{\rm AS}}=\frac{\langle \tilde\Delta_{i+1,t+1}\rangle}{z_c},
$$
where $\langle\cdot\rangle$ denotes the ensemble average due to
the stochasticity in toppling rules. When $0<p_{_{\rm AS}}<1$,
such toppling bias does not change the scaling properties of
avalanche dynamics (numerically confirmed but not shown here).

\subsection{Non-Abelian cases}

In NS DSMs, spatially correlated
metastable patterns are well described as the algebraic decay of
the grain density along the avalanche propagation
direction~\cite{Hughes2002}. Due to such spatial correlations in
metastable patterns, the uncorrelated noise $\eta$ in the
avalanche flow equation for Abelian cases should be replaced by
some correlated noise to represent NS metastable patterns. Naively
speaking, one can interpret $\eta$ in Eq.~(\ref{eq:floweqAS}) as
$\rho_{\rm sc}$ having the same dimension of $t^{-1/2}$ as $\eta$.
Thus, we replace $\rho_{\rm sc}$ by $\rho$ in the NS, while
keeping $\sqrt{w(t)}$ due to the stochasticity in toppling rules:
\begin{equation}
\frac{dN(t)}{dt}\approx\sqrt{w(t)}\rho(t),\label{eq:floweqNS}
\end{equation}
which yields $D_h=1-\frac{3}{2}\alpha$. As a result all other
avalanche exponents in the NS can be written in terms of $\alpha$,
such as $\tau_s=\frac{2(3-\alpha)}{4-\alpha}$ and
$\tau_t=2-\frac{\alpha}{2}$.

The value of $\alpha$ in the NS is numerically observed as $0.45$,
slightly less than $\frac{1}{2}$. In the earlier
work~\cite{JoHa2008}, it is argued that the deviation seems to be
attributed to logarithmic corrections to scaling. So we simply
assume the random walk behaviors of avalanche boundaries, such
that $\alpha=\frac{1}{2}$, irrespective of the toppling bias and
the threshold value (numerically confirmed but not shown here).
Then, all avalanche exponents in the NS becomes exactly the same
as those in the AS: $\tau_s=\frac{10}{7}$, $\tau_t=\frac{7}{4}$,
and $D_h=\frac{1}{4}$. From now on, we call the set of these
exponents as the ``NS" universality class. One can say that the NS
belongs to the same universality class of the AS in the following
sense: for Abelian cases, avalanche flow can sweep and lose many
grains at the same time due to the uniform grain density. On the
other hand, for non-Abelian cases, avalanche flow can sweep only
few grains due to the power-law decaying grain density, but it can
also leave only few grains behind as taking all grains at unstable
sites. For both cases, the scaling property of $N(t)$ is
apparently unaffected by the grain density as long as the toppling
rule is stochastic. Finally, we naively propose another possible
scenario for $\alpha=\frac{1}{2}$ in the NS by mapping metastable
patterns onto the (1+1)-dimensional spatiotemporal configuration
of $2A\to A$ coagulation-diffusion model, where the particle
density temporally decays as $t^{-1/2}$~\cite{Hinrichsen2000}.
However, this scenario requires further investigation.
\begin{table}[t]
\caption{Avalanche exponents $\{\tau_s,\ \tau_t,\ D_h\}$, grain
density and scar exponents, $\alpha$ and $\alpha_{\rm sc}$
and our conjecture in two-dimensional DSMs:
the error bars of numerical results are obtained from effective exponent plots by
the assumption of the dominant simple power-law
scaling~\cite{errors}. It is worthwhile to note that
$D_s=\tau_t$ and $D_t=1$ in all DSMs and $\alpha_{\rm sc}=\alpha$
in non-Abelian DSMs.} \label{table:exponent}
\begin{tabular}{l*{5}{c}}
\hline\hline Model & $\tau_s$ & $\tau_t$ & $D_h$ & $\alpha$,
$\alpha_{\rm sc}$ \\ \hline

Mean field ($d_u=2$) & $\frac{3}{2}$ & $2$ & $0$ & &\\
\\
Abelian\\
$~~$ Deterministic & $\frac{4}{3}$ & $\frac{3}{2}$ & $0$ & $0$,
$\frac{1}{2}$\\

$~~$ Stochastic & $\frac{10}{7}$ & $\frac{7}{4}$ & $\frac{1}{4}$ &
$0$, $\frac{1}{2}$\\
\\
Non-Abelian\\
$~~$ Stochastic & $\frac{2(3-\alpha)}{4-\alpha}$ &
$2-\frac{\alpha}{2}$ & $1-\frac{3\alpha}{2}$ & $\alpha$\\
$~~~~~$ - $z_c=2$ & $1.43(1)$ & $1.78(1)$ & $0.31(3)$ & $0.45(3)$  \\

\\
$~~$ Deterministic & $\frac{3}{2}$ & $2$ & $1-\alpha$ &  $\alpha$\\

$~~~$ aND \\
$~~~~~$ - $z_c=2$ & $1.49(1)$ & $1.94(1)$ & $0.07(1)$ & $0.86(3)$ \\
$~~~~~$ - $z_c=3$ & $1.46(2)$ & $1.78(3)$ & $0.26(1)$ & $0.37(1)$ \\
$~~~~~$ - $z_c=4$ & $1.53(1)$ & $2.00(5)$ & $0.16(2)$ & $0.97(1)$ \\

$~~~$ bND\\
$~~~~~$ - $z_c=2$ & $1.43(1)$ & $1.79(1)$ & $0.06(1)$ & $0.69(5)$ \\
$~~~~~$ - $z_c=3$ & $1.46(2)$ & $1.85(3)$ & $0.20(2)$ & $0.78(14)$ \\
$~~~~~$ - $z_c=4$ & $1.51(2)$ & $2.02(3)$ & $0.15(1)$ & $0.89(15)$ \\

$~~~$ cND\\
$~~~~~$ - $z_c=2$ & $1.52(3)$ & $2.04(3)$ & $0.06(4)$ & $0.91(11)$ \\
\hline\hline
\end{tabular}
\end{table}

In ND DSMs, there are no fluctuations
if we focus on the continuous version, namely the cND, shown in
Fig.~\ref{fig:toppling}(c). The avalanche flow equation can be
written as
\begin{equation}
\frac{dN(t)}{dt}\approx w(t)\rho(t),
\end{equation}
yielding $D_h=1-\alpha$. We find that the scaling exponents of
avalanche mass and duration are MF values, independent of
$\alpha$, i.e., $\tau_s=\frac{3}{2}$ and $\tau_t=2$, whereas other
exponents still depend on $\alpha$. In that sense, we point out
that one should check other avalanche exponents besides those of
the avalanche mass and duration to discuss the universality class
in the ND. If the assumption of $\alpha=D_w=1$ is accepted based
on the linear behavior of avalanche boundaries, all avalanche
exponents turn to be MF values, except for the case of the
avalanche width~\cite{MFexceptw}. Furthermore, this may also
correspond to the MF behavior of the ($d+1$)-dimensional
spatiotemporal configuration of the coagulation-diffusion model
when $d\ge d_u=2$, where the particle density decays as
$t^{-1}$~\cite{Hinrichsen2000}. The peculiar ``MF" behavior in the
two-dimensional ND, i.e., appeared in the low dimensional system,
can be easily understood in the context of the shape of $N(t)$
with its width and height.

Toppling rules we considered in the ND suppress the fluctuations
of the height profile in avalanche flow much more than those in
the NS do. This leads to spread grains wider and wider and makes
avalanche boundaries grow faster, almost ballistically. Such a
positive feedback enables $D_w=1$ to be larger than $\frac{1}{2}$
for all other DSMs. The resultant $D_h=0$ indicates the MF
behavior with $\alpha=1$, whereas $D_h=0$ for any dimensions in
the AD.

Table~\ref{table:exponent} presents the avalanche exponents, the grain
density exponent, and the scar exponent, as well as the exact results
for the Abelian case and our conjecture for the non-Abelian case.

We report that the numerical results in the aND and the bND
clearly deviate from MF results. As the possible origin of
anomalous scaling behaviors, we claim that they are attributed to
the toppling bias applied to the last grain when $z(i,t)=2k+1\geq
z_c$ in Eq.~(\ref{eq:NDrules}). The toppling bias effect in the ND
is numerically tested as $z_c$ increases. We find that the
increment of $z_c$ may yield either that the toppling bias becomes
negligible compared to other ($2k$) grains, or that it turns to be
relevant to change the universality class. The aND with even $z_c$
values and the bND with all $z_c$ values correspond to the former.
They show the MF behavior with some logarithmic corrections for
small $z_c$ values and without such corrections for large $z_c$
values as expected (see Table~\ref{table:exponent}). Scaling
behaviors in the aND with odd $z_c$ values are clearly deviated
from MF results, and seem to be those of the NS class.

Two different mechanisms can be illustrated for the MF class and
the NS class emerged in the aND against the parity of $z_c$. After
the transient period of sandpile models, each site having ever
toppled should be empty or occupied by at least $[{z_c}/2]$
grains. Here $[m]$ is the integer value that is smaller than or
equal to $m$.

In the even case with $z_c=2k$, at least $k$ grains at the
rightmost toppled site of the avalanche, say, $(i,t)$, topple to
its right neighboring site on the next layer, $(i+1,t+1)$. The
site $(i+1,t+1)$ will topple if occupied by grains, or will not if
empty. At small $t$, the grain density is relatively high and the
avalanche width is still small. Grains swept by avalanche flow
quickly diffuse to avalanche boundaries, which causes the
ballistic behavior. On the other hand, at large $t$, remaining
grains are rare in metastable patterns. Avalanche flow keeps
losing grains at avalanche boundaries, which also causes the
ballistic behavior, resulting in $\alpha=1$.

In the odd case with $z_c=2k+1$, the minimal number of grains at a
stable site is $k$ unless empty. Assume that $z_c$ grains are at
the rightmost site of an avalanche, say $(i,t)$, and its right
neighboring site $(i+1,t+1)$ is occupied by $k$ grains. The site
$(i+1,t+1)$ will topple only if it receives $k+1$ grains from the
site $(i,t)$, or will not if it receives $k$ grains. Whether the
site topples or not depends on the value of the toppling arrow
$a(i,t)$. This randomness inherent to toppling arrows leads to the
random walk behaviors of avalanche boundaries, i.e.,
$\alpha=\frac{1}{2}$. If more than $z_c$ grains topple at the
rightmost site or if its right neighboring site is occupied by
more than $k$ grains, the randomness of the alternative toppling
bias is suppressed as in the aND with even $z_c$ values. The same
scaling behavior is also observed in the pND with $p=\frac{1}{2}$,
where the toppling bias is completely random. In the bND with no
intrinsic randomness (the pND with $p=1$), we do not expect the
random walk behaviors of avalanche boundaries, irrespective of the
parity of the threshold value. The pND in the range of
$\frac{1}{2}<p<1$ will be discussed with numerical results in
Sec.~\ref{Sect:numerics}.

\section{Numerical results}
\label{Sect:numerics}

We performed extensive numerical simulations for all DSMs to
confirm our conjecture for avalanche exponents in terms of the
scar exponent, $\alpha_{\rm sc}$ ($\alpha=\alpha_{\rm sc}$ for
non-Abelian cases), up to $L=T=2^{13}$ ($T=2^{15}$ or $L=2T$ in
some cases).

Using conventional successive slope techniques for at most $10^9$
avalanches in the steady state enough after the transient period,
we measure the avalanche exponent set $\{\tau_x,D_x\}$ for all $x$
and the grain density exponent $\alpha$ ($\alpha_{\rm sc}$ for
Abelian cases). As discussed, spatially correlated scars are
observed in all DSMs with the non-zero values of $\alpha_{\rm
sc}$. The spatial correlations of metastable patterns are only
observed in non-Abelian DSMs with the non-zero values of $\alpha$
(see Table~\ref{table:exponent} and
Fig.~\ref{fig:patterns_vs_scars}).

We calculate effective exponents from power-law distributions of
various avalanche quantities, scaling relations among those
quantities, and the algebraic decay of the grain occupation
density in terms of following definitions:
\begin{eqnarray}
\tau_x^{\rm eff}(x)&=&-\frac{\ln P(x)-\ln P(x/m)}{\ln x - \ln
(x/m)},\label{eq:tau_eff}\\
D_x^{\rm eff}(t) &=&\frac{\ln E[x|t]-\ln E[x|t/m]}{\ln t-\ln (t/m)},\\
\alpha^{\rm eff}(t)&=&-\frac{\ln \rho_{_\theta}(t)-\ln
\rho_{_\theta}(t/m)}{\ln t-\ln (t/m)},
\end{eqnarray}
as taking appropriate values of $m$. In order to get the
reasonable resolution of effective exponents, we set $m=5$ for
$\tau^{\rm eff}$ and $\alpha^{\rm eff}$ and $m=1$ for $D^{\rm
eff}$, respectively. The exponents, $\tau_x$ and $D_x$
($=\gamma_{xt}$), are directly obtained by the linear fit of
scaling regimes in power-law distribution functions and scaling
relations. Scaling regimes can be identified either from the flat
region of effective exponents or the systematic tendency as the
system size increases. Regarding $\gamma_{yx}$, instead of the
original definition, $E[y|x]\sim x^{\gamma_{yx}}$, we make use of
the modified definition, $E[y|x]=x+C x^{\gamma_{yx}}$, when $y\geq
x$ for each avalanche~\cite{Chessa1999}. Due to $s\geq z_ct$ and
$h\geq z_c$ by definition, we obtain $\gamma_{st}$ and
$\gamma_{ht}$ using $E[s|t]=z_ct+C t^{\gamma_{st}}$ and
$E[h|t]=z_c+C't^{\gamma_{ht}}$, respectively. The values of
scaling exponents are listed in Table~\ref{table:exponent}, which
confirm our conjecture by means of the avalanche flow analysis in
Sec.~\ref{Sect:flow analysis}.

\begin{figure*}[]
\begin{tabular}{cc}
\includegraphics[width=.8\columnwidth]{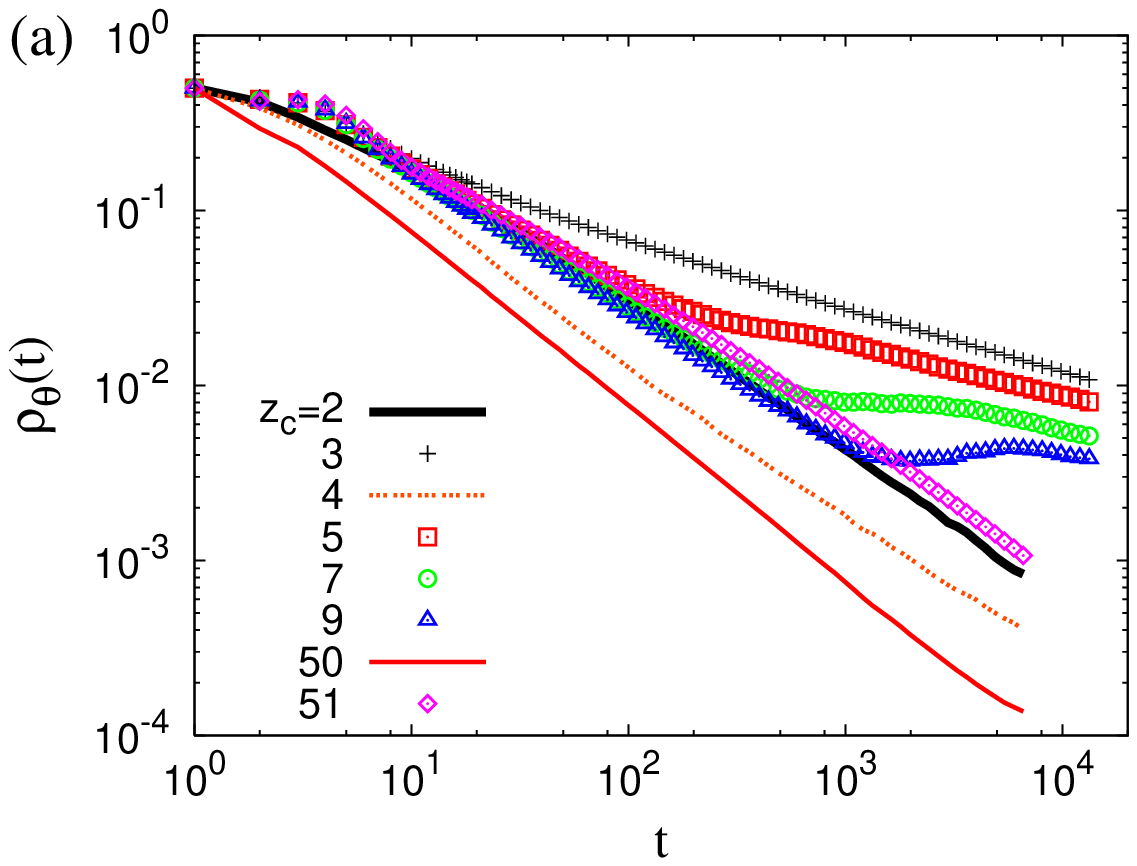}&
\includegraphics[width=.8\columnwidth]{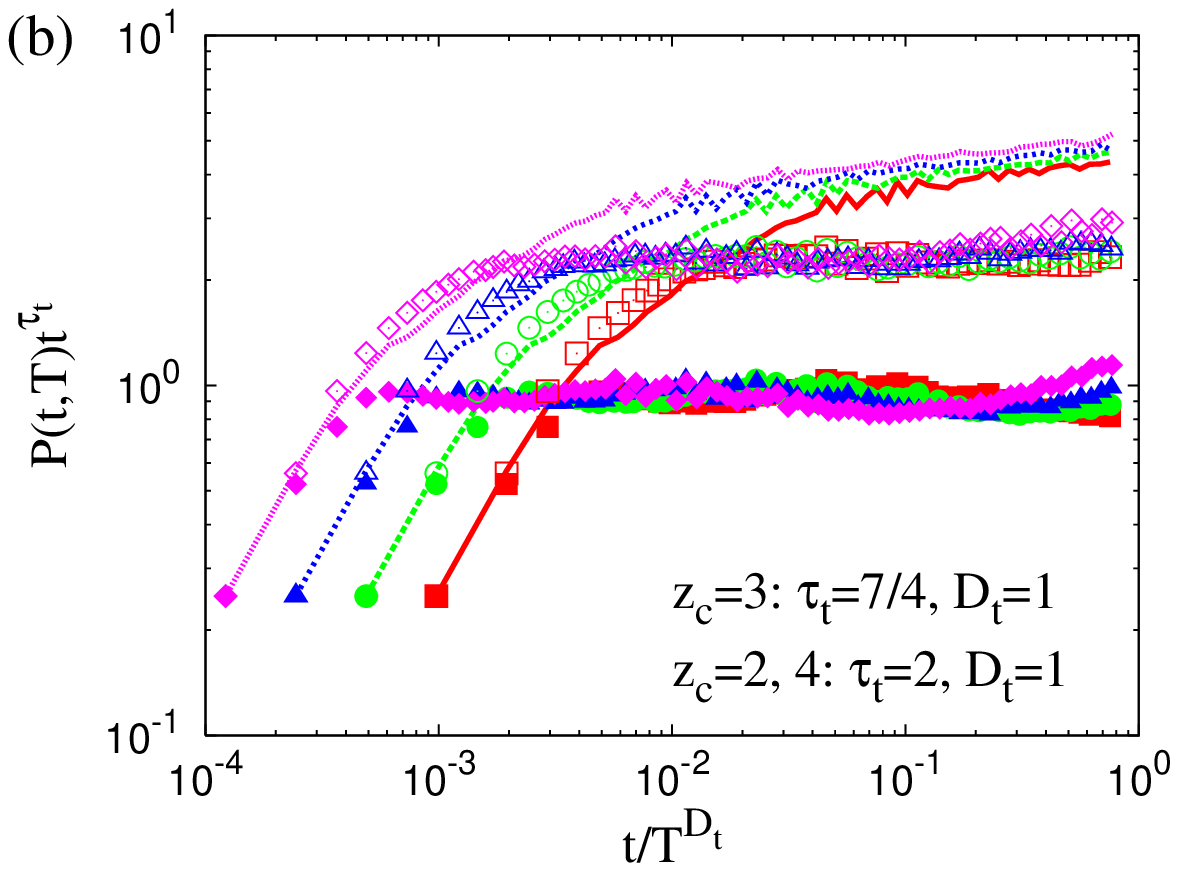}\\
\includegraphics[width=.8\columnwidth]{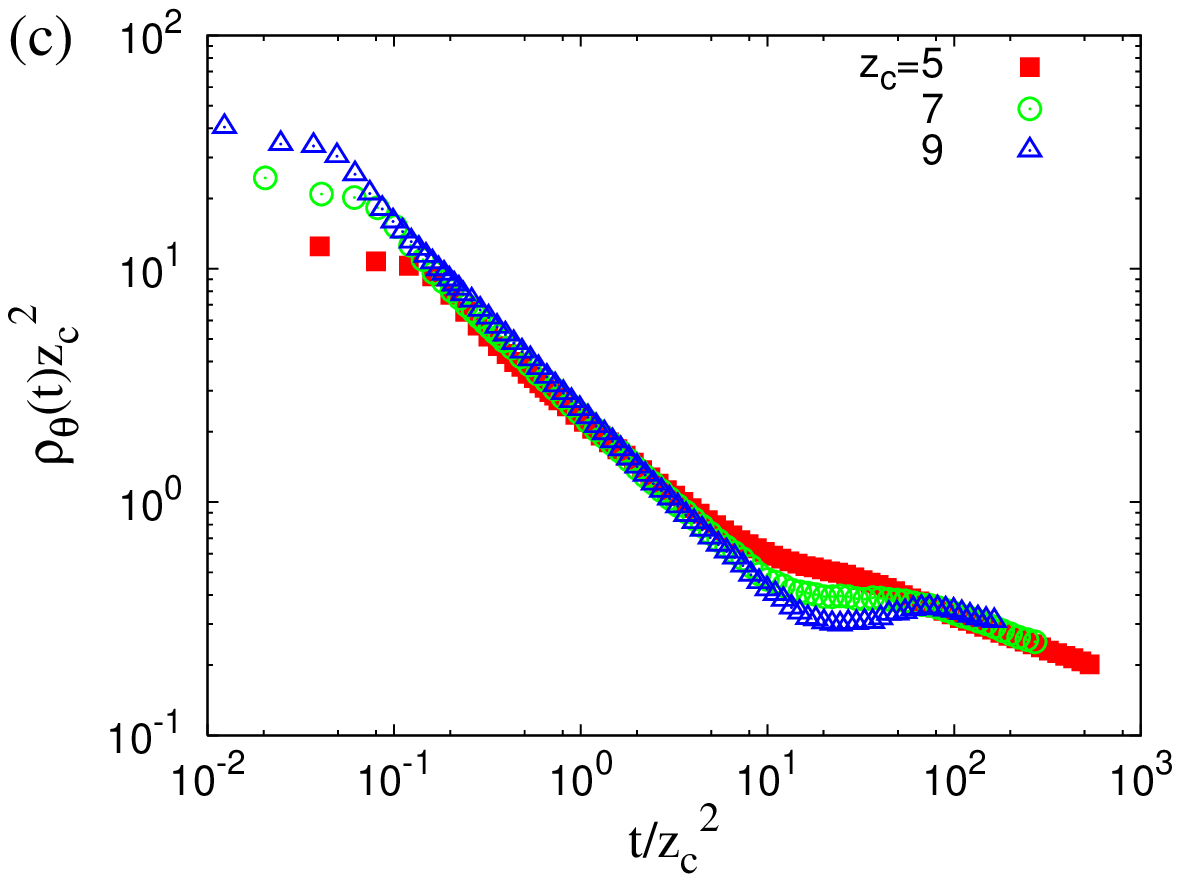}&
\includegraphics[width=.8\columnwidth]{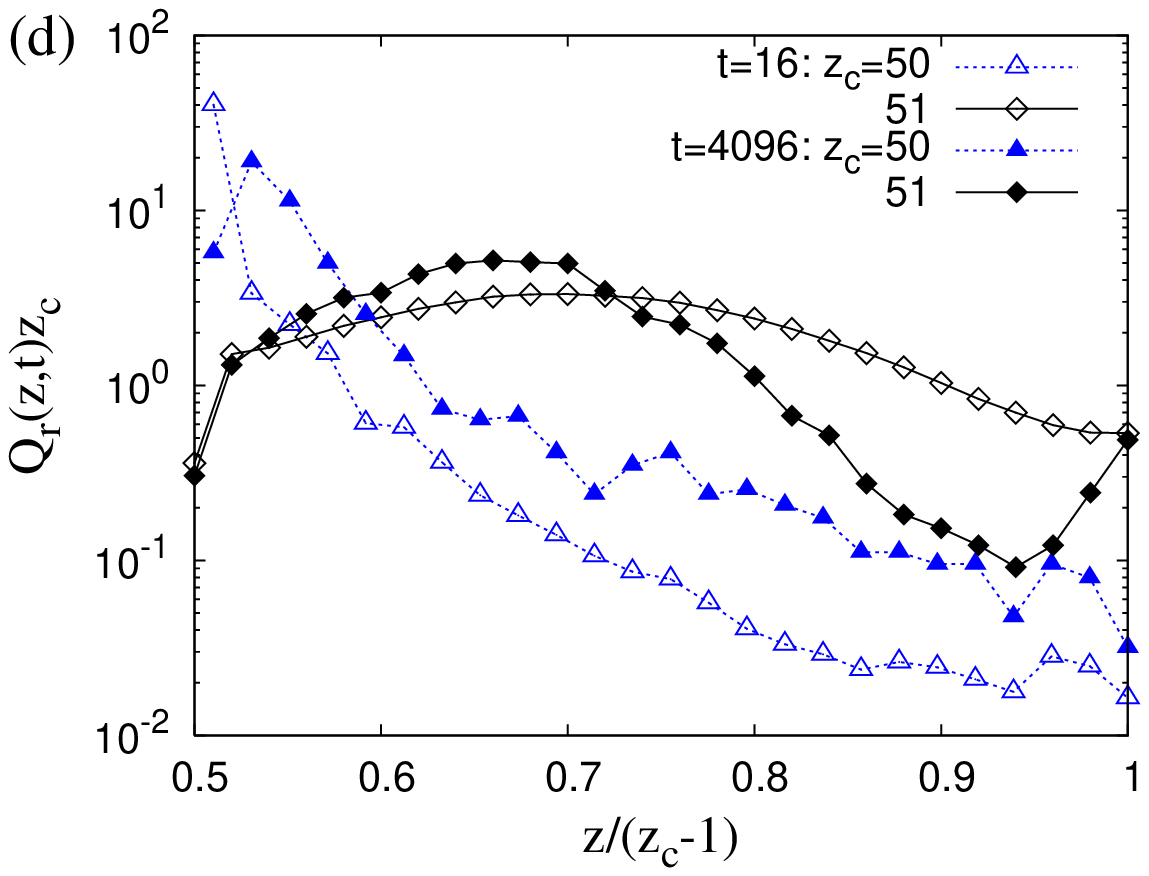}
\end{tabular}
\caption{(Color online) Numerical results of the aND for various $z_c$ values: (a) grain occupation densities $\rho_{_\theta}(t)$ versus $t$ on a lattice with $T=2^{14}$ for $z_c=3$, $5$, $7$, and $9$ ($T=2^{13}$ otherwise) and $L=2T$. (b) Finite-size scaling of $P(t,T)$ with the MF value ($\tau_t=2$) for $z_c=2$ (lines) and $z_c=4$ (open symbols), and with the NS value ($\tau_t=7/4$) at $z_c=3$ (filled
symbols). Here, $T=L=2^{10}$, $2^{11}$, $2^{12}$, and
$2^{13}$ from right to left and $D_t=1$ for all cases. (c) Crossover scaling of grain occupation densities for odd $z_c$ values by $\rho_{_\theta}(t)z_c^2$ as a function of $t/z_c^2$, which are the same data as (a). (d) Rescaled grain distributions $Q_r(z,t)$ at two different layers, $t=16$ (open symbols) and $4096$ (filled symbols) for $z_c=50$ ($\triangle$) and $51$ ($\diamond$) on a lattice with $T=2^{13}$ and $L=2T$.}
\label{fig:aND}
\end{figure*}
\begin{figure*}[]
\begin{tabular}{cc}
\includegraphics[width=.8\columnwidth]{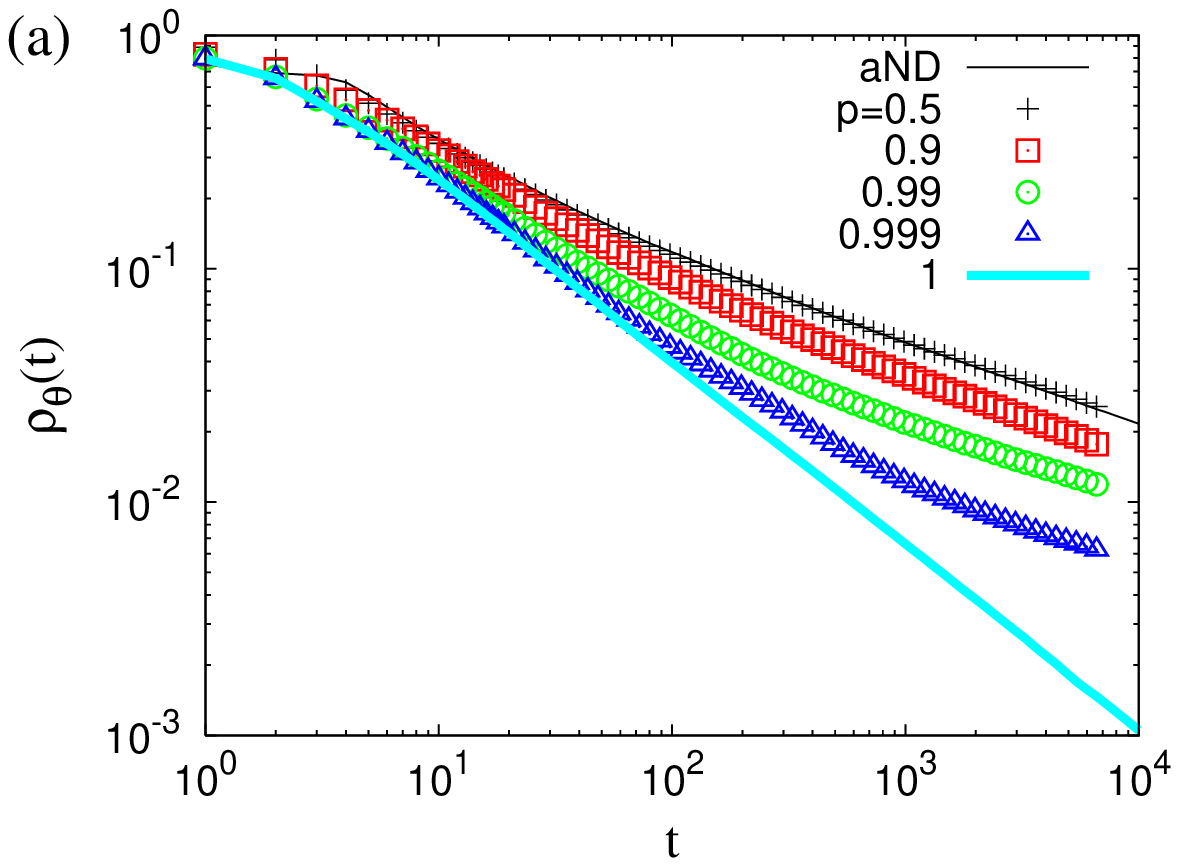}&
\includegraphics[width=.8\columnwidth]{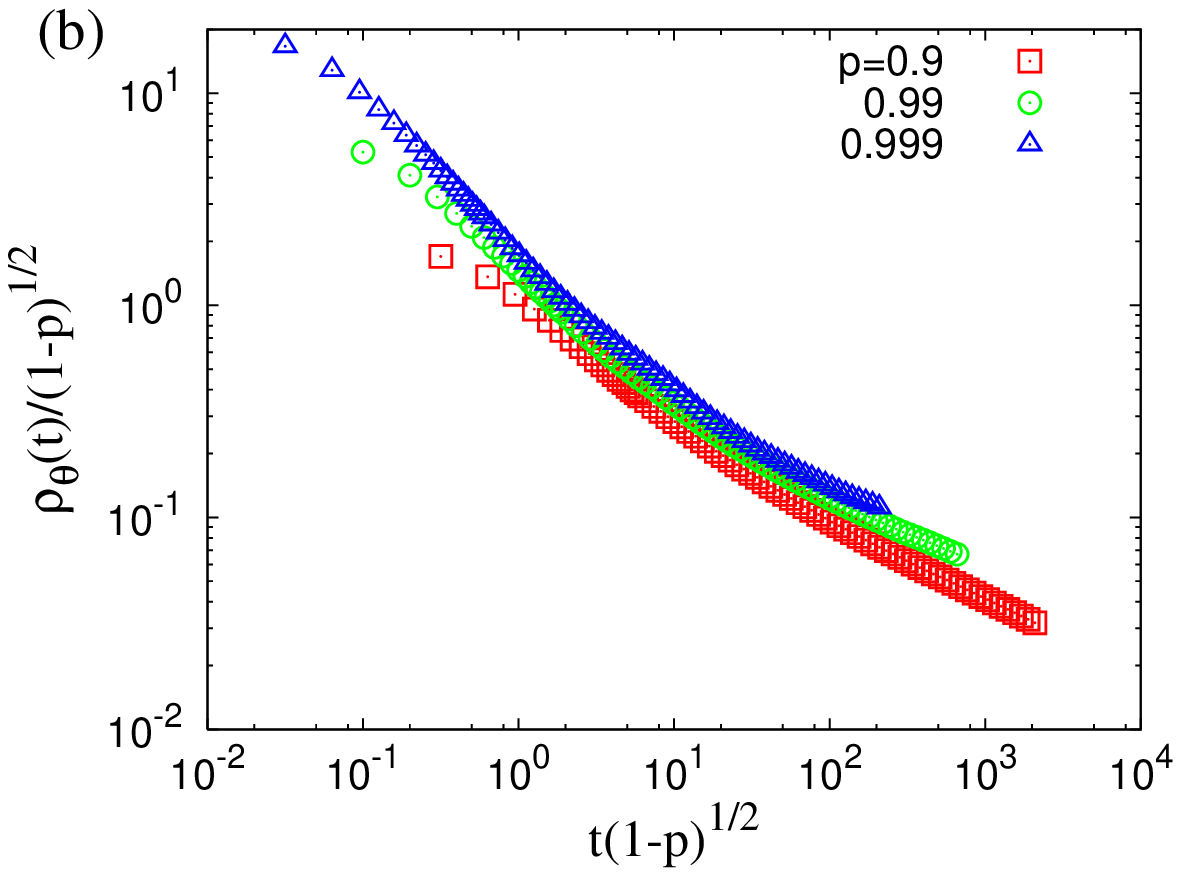}
\end{tabular}
\caption{(Color online) (a) Grain occupation densities $\rho_{_\theta}(t)$ versus $t$ in the pND at a fixed $z_c=3$ as $p$ increases from $0.5$ (the aND as the thin line at the top) to $1$ (the bND as the thick line at the bottom). We used a lattice with $T=2^{13}$ and $L=2T$.
(b) Crossover scaling of grain occupation densities at $z_c=3$ by
$\rho_{_\theta}(t)/(1-p)^{1/2}$ as a function of $t(1-p)^{1/2}$.}
\label{fig:pND_rho}
\end{figure*}
\begin{figure}[]
\includegraphics[width=.8\columnwidth]{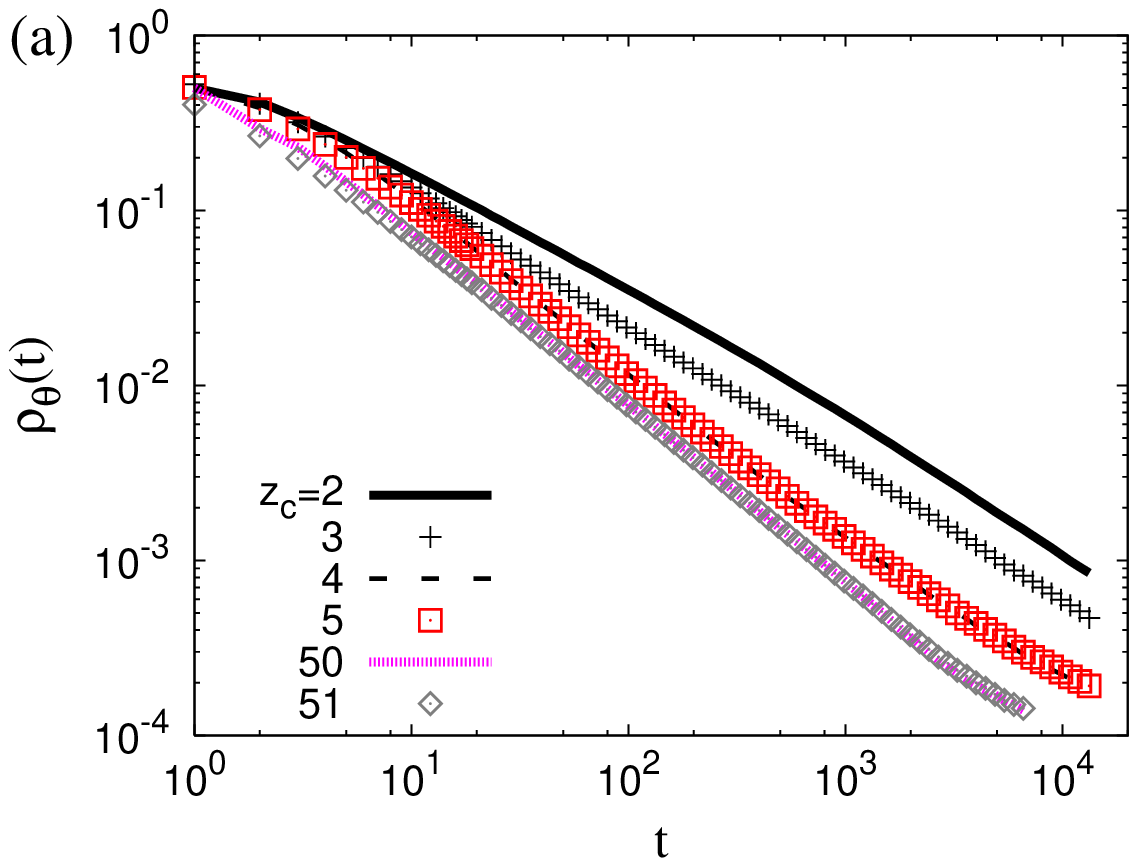}\\
\includegraphics[width=.8\columnwidth]{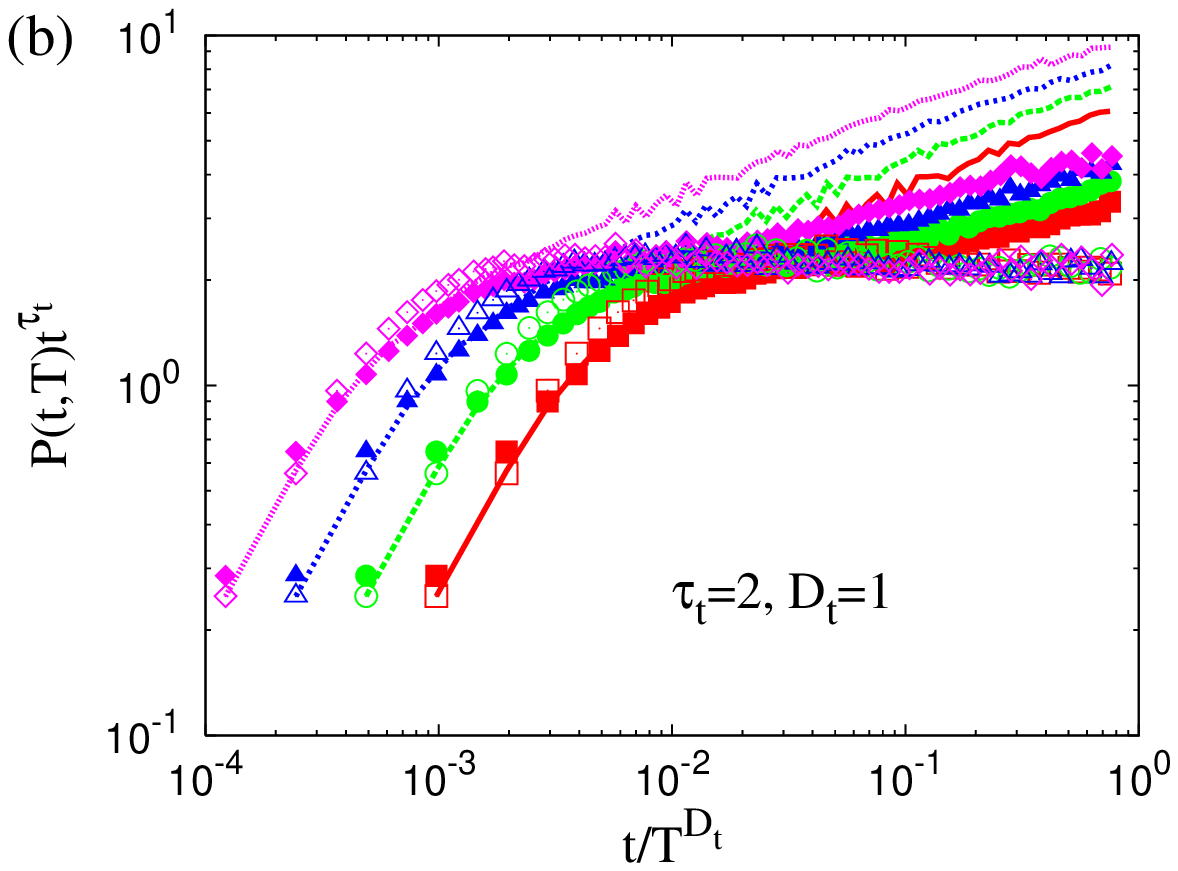}\\
\includegraphics[width=.8\columnwidth]{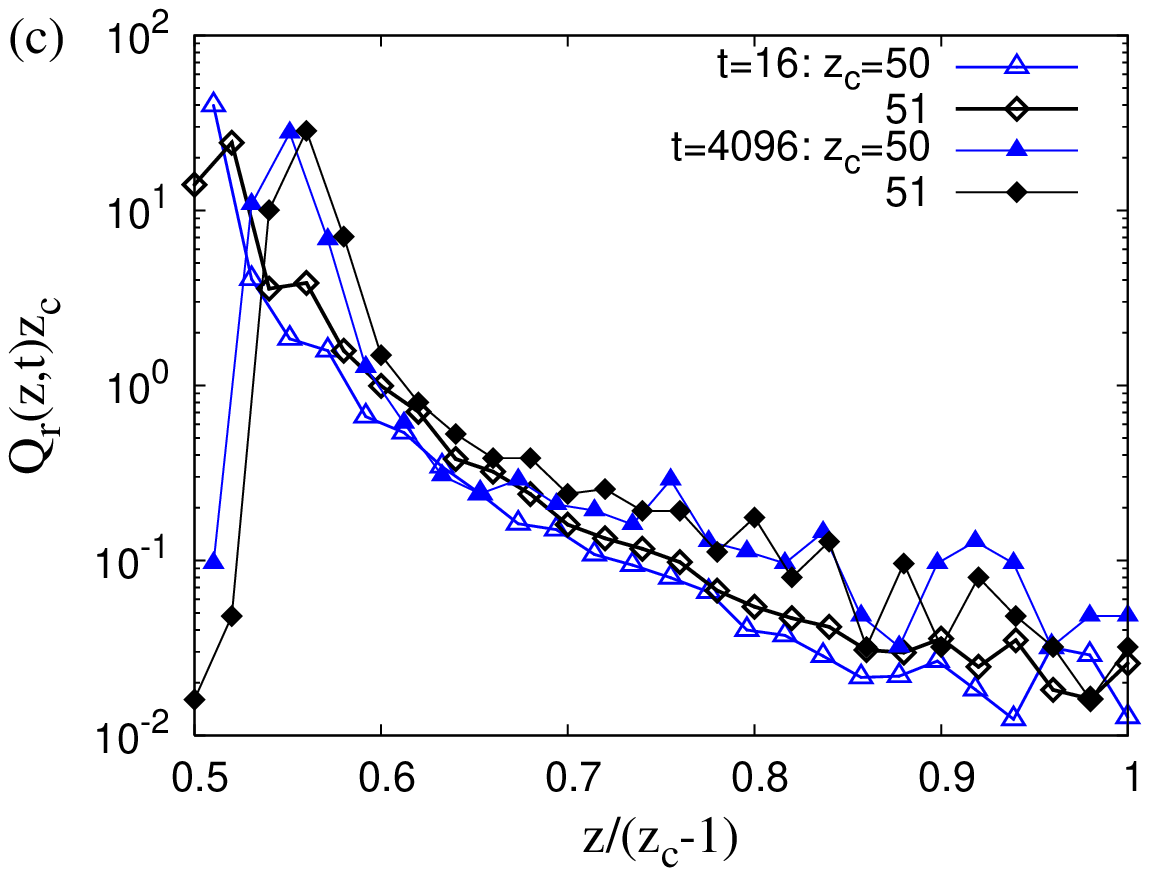}
\caption{(Color online) Numerical results of the bND for various $z_c$ values: (a) grain occupation densities $\rho_{_\theta}(t)$ versus $t$ on a lattice with $T=2^{14}$ and $L=2T$ ($T=2^{13}$ for $z_c=50$ and $51$), where lines for $z_c=2$, $4$, and $50$ are from top to bottom. (b) Finite-size
scaling of $P(t,T)$ with MF values for $z_c=2$ (lines), $z_c=3$
(filled symbols), and $z_c=4$ (open symbols). Here $T=L=2^{10}$, $2^{11}$, $2^{12}$, and $2^{13}$ from right to left. (c) Rescaled grain distributions $Q_r(z,t)$ at two different layers, $t=16$ (open symbols) and $4096$ (filled symbols) for $z_c=50$ ($\triangle$) and $51$ ($\diamond$) on a lattice with $T=2^{13}$ and $L=2T$.} \label{fig:bND}
\end{figure}
\begin{figure}[]
\includegraphics[width=.8\columnwidth]{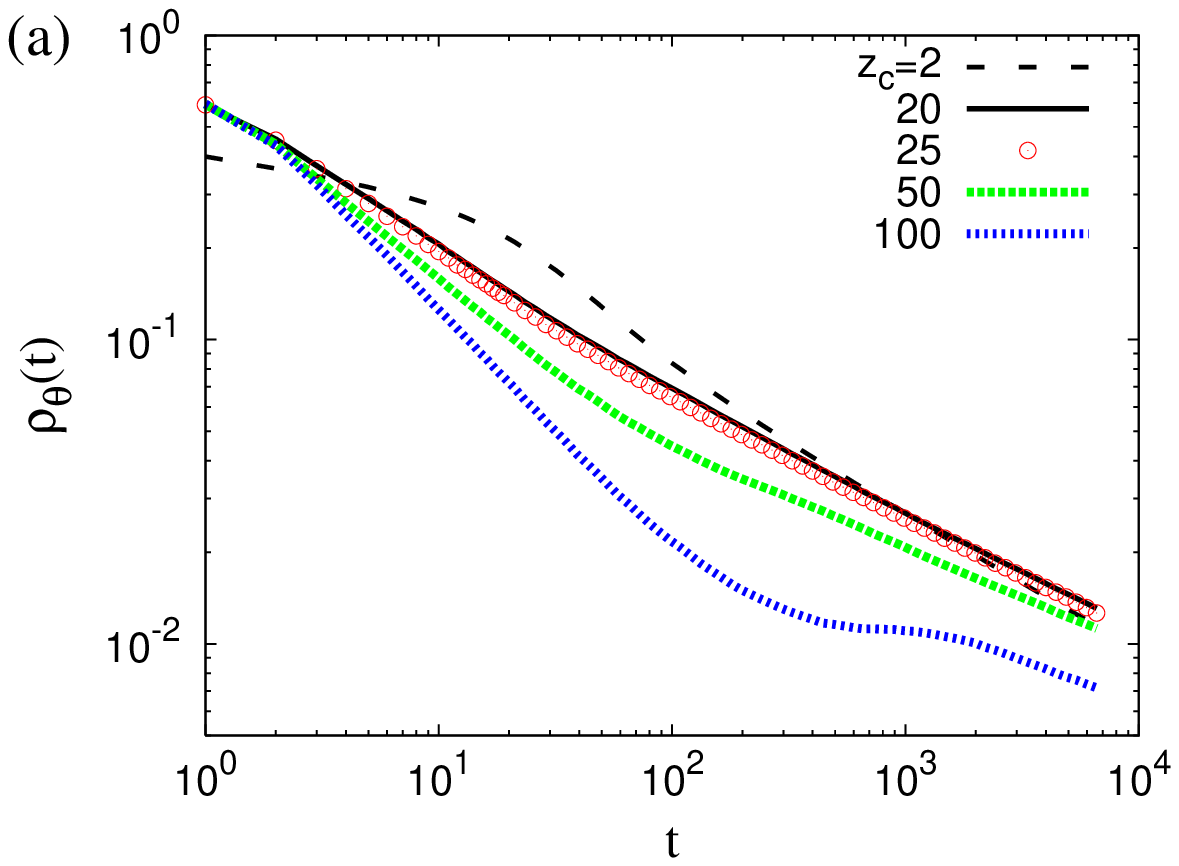}\\
\includegraphics[width=.8\columnwidth]{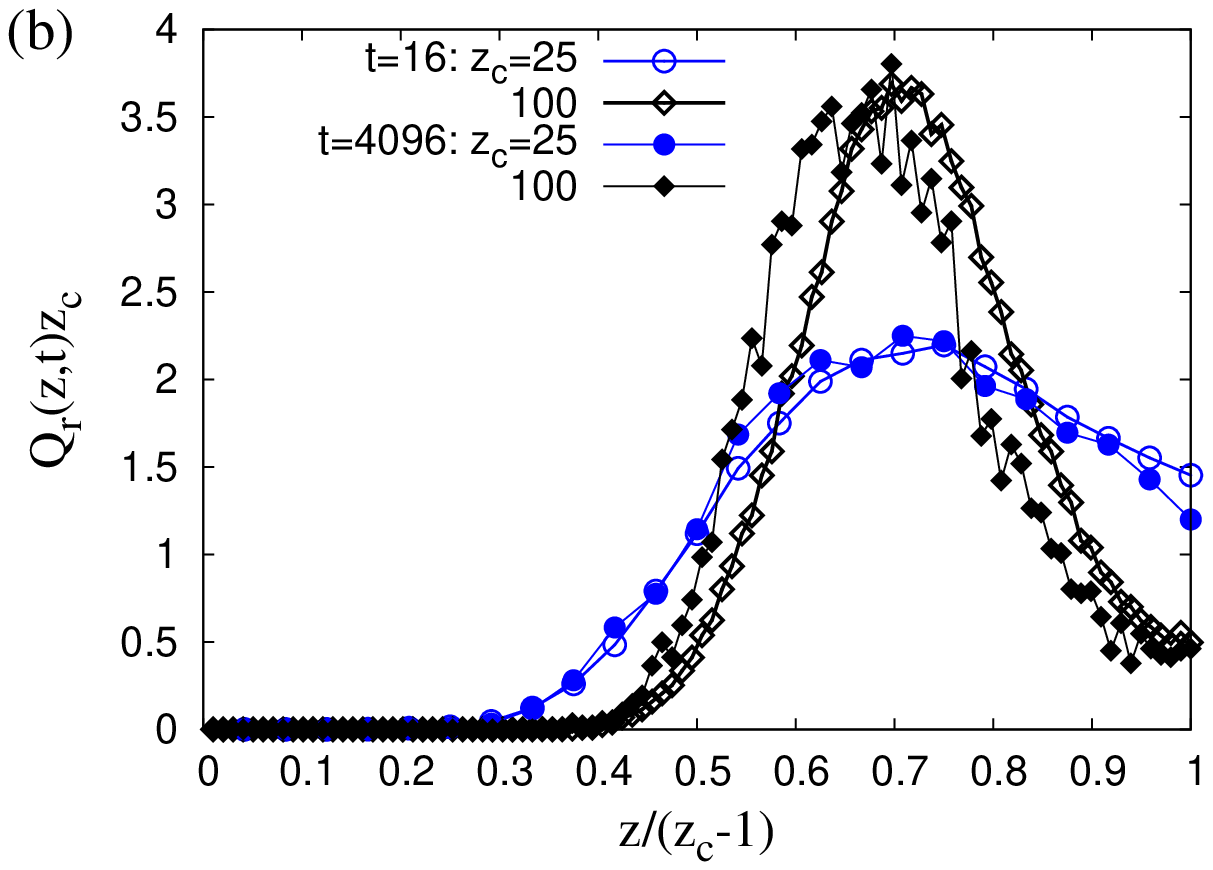}
\caption{(Color online) (a) Grain occupation densities $\rho_{_\theta}(t)$ versus $t$ in the NS for various $z_c$ values (lines for $z_c=2$, $20$, $50$, and $100$ from top to bottom at the right). We used a lattice with $T=2^{13}$ and $L=2T$. (b) In particular to the cases for $z_c=25$ and $100$, rescaled grain distributions, $Q_r(z,t)$, are measured at two different
layers, $t=16$ (open symbols) and $4096$ (filled symbols).} \label{fig:NS_all}
\end{figure}

In the aND, the scaling behaviors of grain occupation densities
definitely depend on the parity of the threshold value, $z_c$, as
shown in Fig.~\ref{fig:aND}(a). For the even $z_c$ values, the
scaling regime of $\alpha^{\rm eff}(t)$ approaches $1$ as $z_c$
increases (not shown). The value of $\alpha$ at $z_c=3$ appears to be
$\frac{1}{2}$ within errors rather than $1$. Interestingly, for
the larger odd $z_c$ values, we observe the crossover behavior
from $\alpha^{\rm eff}(t)=1$ (with logarithmic corrections) to
$\frac{1}{2}$ at some ``crossover layer,'' $t_\times$, which
clearly increases as $z_c$ increases. We investigate the same
parity effect on the universality class in terms of other
effective exponents, $\tau_s^{\rm eff}$, $\tau_t^{\rm eff}$, and
$D_h^{\rm eff}$, for avalanche distributions (not shown).
The effective exponents for $z_c=2$ and $4$ approach MF values
($\tau_s=\frac{3}{2}$, $\tau_t=2$, and $D_h=0$) or
at least have the tendency to approach them.
On the other hand, effective exponents at $z_c=3$ clearly
deviate from MF values. We also confirm the finite-size scaling (FSS) behaviors
of avalanche distributions with MF values for $z_c=2$ and $4$, and
with NS values ($\tau_s=\frac{10}{7}$, $\tau_t=\frac{7}{4}$, and
$D_h=\frac{1}{4}$) at $z_c=3$. Figure~\ref{fig:aND}(b) shows
the successful FSS collapses of avalanche duration distributions.

Numerical data in the aND are averaged over a lot of avalanches
(up to $10^9$ with only few numbers of ensemble for random initial
configuration setups) in the steady state after the transient.
Since toppling arrows, $\{a(i,t)\}$, are annealed enough by the
random addition of grains in the steady state, it is guaranteed
that all the results are completely independent of initial
conditions, which are also numerically confirmed.

Figure~\ref{fig:aND}(c) shows the crossover scaling of
grain occupation densities in the aND for odd $z_c$ values. It can
be written as
\begin{equation}
\rho_{_\theta}(t)=t^{-1}g(t/t_\times),
\end{equation}
where the scaling function behaves as $g(x)\sim \mathcal{O}(1)$
for small $x$ and $g(x)\sim \sqrt{x}$ for large $x$. At small $x$
($t<t_\times$), grain occupation densities are relatively high
while the avalanche width is still small. Grains at avalanche bulk
sites quickly diffuse to avalanche boundaries. The numbers of
grains at avalanche boundaries tend to be larger than $z_c$ and
thus make ballistic avalanche boundaries ($\alpha=1$). As the
grain occupation densities decay to become low at large $x$
($t>t_\times$), the numbers of grains at avalanche boundaries tend
to be $z_c$. Then the stochasticity of avalanche boundaries is
disclosed ($\alpha=\frac{1}{2}$). The value of $t_\times$
indicates the crossover from the dense region to the sparse region
of remaining grains. We present some naive test for the crossover
scaling of $\rho_{_\theta}$ using $t_\times\sim z_c^2$. Indeed,
numerical data collapse pretty well for $z_c=5$, $7$, and $9$, as
shown in Fig.~\ref{fig:aND}(c).
This can be explained by an intuitive argument.
As discussed in the last part of Sect.~\ref{Sect:flow analysis},
the stochasticity inherent to toppling arrows reveals
only in the case when the unstable boundary site,
say, $(i,t)$ at the right boundary, has $z_c$ grains
and its neighboring site at the next layer, say, $(i+1,t+1)$, is
occupied by $k$ grains. Assuming that the grain distribution is
uniform (to be confirmed), the probability of an occupied site having $k$ grains is
about $z_c^{-1}$. If the unstable site has $z_c$ grains, it might
have been occupied by $k$ or $k+1$ grains before receiving grains
from the unstable neighboring site at the previous layer, whose
probability is about $z_c^{-1}$. Then, the stochasticity in
toppling arrows begins to appear when the expected number of such
cases up to the layer $t$ becomes finite, i.e., $z_c^{-2}t\sim
\mathcal{O}(1)$, which implies that $t_\times\sim z_c^2$.

In the aND at $z_c=51$, the value of $\alpha$
approaches $1$ in Fig.~\ref{fig:aND}(a), which may mislead to
conclude no crossover for the larger $z_c$ values even if they are
odd. We here pose the following question: {\it How can we know
whether or not any crossover behavior is observed at the
sufficiently large layer?} The value of $\alpha$ is not sufficient
enough to answer the question. We need to investigate the grain
distribution $Q(z,t)$ in Eq.~(\ref{eq:Qzt}) that contains the
detailed information for metastable grain patterns. Due to the fact
that the $Q(0,t)=1-\rho_{_\theta}(t)$ becomes dominant as $t$
increases for non-Abelian DSMs, we consider the rescaled grain
distribution as
\begin{equation}
Q_r(z,t)=\frac{Q(z,t)}{1-Q(0,t)},
\end{equation}
where $Q_r(z,t)=0$ for $z\in\{1,2,...,[{z_c}/2]-1\}$.

We numerically find that $Q_r(z,t)$ is broad due to the stochastic
nature for odd $z_c$ values. It is also found that it is peaked
around at the value of $[{z_c}/2]$ due to the deterministic
nature for even $z_c$ values. Figure~\ref{fig:aND}(d) shows
the broadness of $Q_r(z,t)$ for $z_c=51$ and the peaked structure
of $Q_r(z,t)$ for $z_c=50$. This implies that
the shape of the grain distribution is one of the good indicators to
identify the universality class, as well as the grain density
exponent in metastable patterns.

The effect of the quenched randomness in the aND is compared to
that of the annealed randomness in the pND.
The scaling behaviors of $\rho_{_\theta}(t)$ in the pND with $p=\frac{1}{2}$
for various $z_c$ values exhibit qualitatively the same behaviors
as those in the aND (not shown). Since the scaling behavior of $\rho_{_\theta}(t)$
in the aND ($\alpha=\frac{1}{2}$) at $z_c=3$ is quite different
from that in the bND ($\alpha=1$) at the same $z_c$ value,
one can ask if another crossover behavior exists
between them. At each value of $p$ close to $1$ in the pND, we are
able to observe the crossover behavior of the grain occupation
density from $\alpha=1$ to $\frac{1}{2}$ at some crossover layer,
$t_\times$, as shown in Fig.~\ref{fig:pND_rho}(a). We numerically
find that the value of $t_\times$ scales as $(1-p)^{-1/2}$, [see
Fig.~\ref{fig:pND_rho}(b)]. Such scaling is roughly explained by
the following argument: the degree of stochasticity per toppling
event is simply given by $1-p$. The total number of realizations
up to the layer $t$ is on the order of $w(t)t\sim t^2$ for
$t<t_\times$. Thus, the stochasticity begins to appear when
$(1-p)t^2\sim \mathcal{O}(1)$, which implies that $t_\times\sim (1-p)^{-1/2}$.

Figure~\ref{fig:bND}(a) shows the scaling behaviors of
$\rho_{_\theta}(t)$ in the bND for various $z_c$ values. The
results imply the MF class with some logarithmic corrections to scaling
for small $z_c$ values and without corrections for large $z_c$ values.
For $z_c=2$ and $3$, $\alpha^{\rm eff}(t)$ values are less than
$1$ even in the scaling regime (not shown), but the possibility to approach
$1$ cannot be excluded. Such deviation affects other scaling behaviors.
We also measure the effective exponents of avalanche distributions
and test the FSS collapse with MF values for $z_c=2$, $3$, and $4$
[see Fig.~\ref{fig:bND}(b)]. From our numerical observation,
it is difficult to determine whether the cases of $z_c=2$ and 3
obey eventually the same FSS as the cases of $z_c\ge 4$ or not.
Figure~\ref{fig:bND}(c) shows that $Q_r(z,t)$ is peaked around at
$[{z_c}/2]$, consistent with the observation that avalanche
boundaries behave ballistically, which implies again the MF class,
independent of the parity of $z_c(\ge 4)$.

Next, we investigate another kind of crossover behaviors from the
NS class to the MF class in the NS as $z_c$ increases up to $100$.
This is motivated by L\"{u}beck's work~\cite{Lubeck2000b} that
reported the crossover behavior between the undirected NS and the
undirected cND for very large $z_c$ values (up to the order of
$10^3$). In the undirected NS, if $m\geq z_c$ grains topple at an
unstable site, its neighboring sites will receive grains of
$\mathcal{O}(m)$ on average with the standard deviation of
$\mathcal{O}(\sqrt{m})$. Thus, for sufficiently large $z_c$
values, the stochasticity in toppling rules is effectively
suppressed and the system shows the deterministic universality
class. It was claimed that universality classes can be determined
in the context of the shape of grain distributions: the grain
distribution is broad for the undirected stochastic class, while
it is narrow with multiple peaks for the undirected deterministic
class. In the similar sense, one can expect essentially the same
crossover from the NS class to the MF class in the directed case
as $z_c$ increases to the sufficiently large value, i.e., by
suppressing fluctuations in $\tilde\Delta_{i\pm 1,t+1}$.

In Fig.~\ref{fig:NS_all}(a), we observe two kinds of
crossover behaviors of grain occupation densities in DSMs. One is
from $\alpha=\frac{1}{2}$ (NS class) to $1$ (MF class) as $z_c$
increases, as expected. The other is from $\alpha=1$ to
$\frac{1}{2}$ at a fixed $z_c$ value as $t$ increases. Here the
crossover layer, $t_\times$, depends on the $z_c$ value. We
systematically observe that the shape of $Q_r(z,t)$ at $z_c=100$
is much narrower than that at $z_c=25$ for any layer $t$ [see
Fig.~\ref{fig:NS_all}(b)]. It implies the MF class for much larger
$z_c$ values, which is hardly confirmed in numerical simulations.

Finally, we discuss the crossover mechanism from the MF class to
the NS class at a fixed $z_c$ value for the directed case. The
mechanism is different from that for the undirected case. The
numbers of toppled grains at unstable sites on the top layer $t=0$
are always $z_c$. Then, $Q(z,t=1)$ has the normal distribution
with the mean value of ${z_c}/2$ and the standard deviation
on the order of $\sqrt{z_c}$, as well as an additional peak around
$z_c-1$. The latter peak is due to the sites that have received
grains both from the neighboring sites on the top layer, but are
still stable. This peaked structure implies the MF class.
Inevitable fluctuations around two peaks in the $Q(z,t)$ at small
$t$ propagate and accumulate through layers, and finally result in
the broad distribution without any distinguished peaks. If the
value of $z_c$ was sufficiently large enough to exclude any
fluctuations (approximately the cND), the MF behavior would be
clearly observed without any crossover behaviors. In other words,
for the finite $z_c$ values, there always exists crossovers from
the MF class to the NS class at some finite crossover layer.

\section{Summary}
\label{Sect:summary}

We studied both non-Abelian stochastic (NS) and mean-field (MF)
universality classes and crossover behaviors between two classes
in various non-Abelian directed sandpile models (DSMs) on a
two-dimensional tilted lattice. The broken Abelian symmetry
induces spatially long-range correlations in metastable patterns.
The universality class of avalanche dynamics can be identified by
means of the grain density exponent in metastable patterns:
$\alpha=\frac{1}{2}$ for the NS class and $\alpha=1$ for the MF
class. Due to some ambiguity of the non-Abelian
deterministic toppling rule against the given lattice structure,
we considered the toppling bias for the last grain. We found that
the parity of the threshold value, $z_c$, might change the
universality class. Using extensive numerical simulations, we
confirmed that MF behaviors are well observed in the aND for even
$z_c$ values and in the bND for any $z_c$ values as $z_c$
increases. For odd $z_c$ values, the toppling randomness in the
aND and in the pND with $p=\frac{1}{2}$ becomes relevant, so that
they belong to the NS class.

In the aND and the pND with $p=\frac{1}{2}$ for large odd $z_c$
values, we numerically observed crossover behaviors from
$\alpha=1$ to $\frac{1}{2}$ at some crossover layer, $t_\times$.
The crossover layer is a function of the $z_c$ value and the
partial bias, $p$, in the pND. Since the grain density exponent is
not sufficient to determine the universality class under some
circumstances, we suggest checking the broadness of the grain
distribution in metastable patterns as well. Here the broadness
originates from the stochastic nature in toppling rules. As in the
undirected case, we also observed that the NS class undergoes the
crossover to the MF class as $z_c$ increases. At a fixed $z_c$
value, another crossover is found from the MF class to the NS
class at a finite value of $t_\times$.

In conclusion, we emphasize that one should be very careful to set
the toppling bias and the threshold value for some appropriate
minimal model of self-organized criticality in order to explain
experimental results. It is because the naive setting of
parameters may mislead to either unexpected universality classes
or crossover behaviors. Finally, it would be worthwhile to extend
our various tests towards some other lattice structures, to
clarify the crossover scaling issues more analytically, and to
test other types of non-Abelian DSMs, such as sticky sandpiles and
sandpiles with bulk dissipation~\cite{DDhar_Comment}.

\vspace*{0.1cm}
\section*{ACKNOWLEDGMENTS}

This work was supported by the BK21 project and by NAP of Korea
Research Council of Fundamental Science and Technology (MH). We
would like to thank D. Dhar for reading our paper and giving
valuable comments on it as well as suggesting further works.



\begin{thebibliography}{00}
\bibitem{CM2005}
K. Christensen and N.R. Moloney, {\it Complexity and Criticality}
(Imperial College Press, London, 2005).

\bibitem{Newman2005}
M.E.J. Newman, Contemp. Phys. \textbf{46}, 323~(2005).

\bibitem{UMM1995}
J.S. Urbach, R.C. Madison, and J.T. Markert, Phys. Rev. Lett.
\textbf{75}, 276~(1995); K.-S. Ryu, H. Akinaga, and S.-C. Shin,
Nat. Phys. \textbf{3}, 547~(2007).

\bibitem{Bak1987}
P. Bak, C. Tang, and K. Wiesenfeld, Phys. Rev. Lett. \textbf{59},
381~(1987).

\bibitem{Jensen1998}
H.J. Jensen, {\it Self-Organized Criticality} (Cambridge
University Press, Cambridge, 1998).

\bibitem{Dhar2006}
D. Dhar, Physica A \textbf{369}, 29~(2006).

\bibitem{YCZhang1989}
Y.-C. Zhang, Phys. Rev. Lett. \textbf{63}, 470~(1989).

\bibitem{Manna1991}
S.S. Manna, J. Phys. A \textbf{24}, L363~(1991).

\bibitem{Ben-Hur1996}
A. Ben-Hur and O. Biham, Phys. Rev. E \textbf{53}, R1317~(1996).

\bibitem{Lubeck1997a}
S. L\"{u}beck and K.D. Usadel, Phys. Rev. E \textbf{55}, 4095~(1997).

\bibitem{Lubeck1997b}
S. L\"{u}beck, Phys. Rev. E \textbf{56}, 1590~(1997).

\bibitem{Milshtein1998}
E. Milshtein, O. Biham, and S. Solomon, Phys. Rev. E \textbf{58},
303~(1998).

\bibitem{Dickman2003}
R. Dickman and J.M.M. Campelo, Phys. Rev. E \textbf{67},
066111~(2003).

\bibitem{Dhar1989}
D. Dhar and R. Ramaswamy, Phys. Rev. Lett. \textbf{63},
1659~(1989).

\bibitem{Pastor-Satorras2000}
R. Pastor-Satorras and A. Vespignani, J. Phys. A \textbf{33},
L33~(2000); Phys. Rev. E \textbf{62}, 6195~(2000).

\bibitem{Paczuski2000}
M. Paczuski and K.E. Bassler, Phys. Rev. E \textbf{62},
5347~(2000); cond-mat/0005340v2.

\bibitem{Kloster2001}
M. Kloster, S. Maslov, and C. Tang, Phys. Rev. E \textbf{63},
026111~(2001).

\bibitem{Hughes2002}
D. Hughes and M. Paczuski, Phys. Rev. Lett. \textbf{88},
054302~(2002); cond-mat/0105408v1.

\bibitem{Pan2005}
G.-J. Pan {\it et al.}, Phys. Lett. A \textbf{338}, 163~(2005).

\bibitem{JoHa2008}
H.-H. Jo and M. Ha, Phys. Rev. Lett. \textbf{101}, 218001~(2008).

\bibitem{Lubeck2000b}
S. L\"{u}beck, Phys. Rev. E \textbf{62}, 6149~(2000).

\bibitem{Hinrichsen2000}
H. Hinrichsen, Adv. Phys. \textbf{49}, 815~(2000).

\bibitem{errors}
For some cases, the deviation from our conjecture may be attributed to
logarithmic corrections to scaling as well as the finite-size effect.

\bibitem{MFexceptw}
In general, on a ($d+1$)-dimensional lattice, $a\sim tw^d$
yielding $D_a=1+dD_w$. In case of $d\geq d_u$, where $d_u=2$ is
the upper critical dimension of DSMs, $dD_w=1$. This is consistent
with our MF results, i.e., $D_w=1$, $d=1$.

\bibitem{Chessa1999}
A. Chessa, H.E. Stanley, A. Vespignani, and S. Zapperi, Phys. Rev. E
\textbf{59}, R12~(1999).

\bibitem{DDhar_Comment}
D. Dhar (private communication).

\end{thebibliography}

\end{document}